\documentclass[%
 aip,
 amsmath,amssymb,
 reprint,%
]{revtex4-1}

\usepackage{graphicx}
\usepackage{dcolumn}
\usepackage{bm}

\usepackage[utf8]{inputenc}
\usepackage[T1]{fontenc}
\usepackage{mathptmx}
\usepackage{etoolbox}
\usepackage{svg}
\usepackage{natbib}
\RequirePackage{silence}
\makeatletter
\def\@email#1#2{%
 \endgroup
 \patchcmd{\titleblock@produce}
  {\frontmatter@RRAPformat}
  {\frontmatter@RRAPformat{\produce@RRAP{*#1\href{mailto:#2}{#2}}}\frontmatter@RRAPformat}
  {}{}
}%
\makeatother
\begin{document}

\preprint{AIP/123-QED}
\title{Submicrometer tunnel ferromagnetic Josephson junctions with transmon energy scale}
\author{R. Satariano}
  \email{roberta.satariano@unina.it}
\affiliation{ 
 Dipartimento di Fisica Ettore Pancini, Università degli Studi di Napoli Federico II, Via Cinthia, 80126, Napoli, IT.
}%

\author{R. Ferraiuolo}
\affiliation{%
QuantWare, Elektronicaweg 10, 2628 XG Delft, The Netherlands
}%

\author{F. Calloni}
\affiliation{ 
 Dipartimento di Fisica Ettore Pancini, Università degli Studi di Napoli Federico II, Via Cinthia, 80126, Napoli, IT.
}%

\author{H. G. Ahmad}
\affiliation{ 
 Dipartimento di Fisica Ettore Pancini, Università degli Studi di Napoli Federico II, Via Cinthia, 80126, Napoli, IT.
}%

\author{D.  Gatta}
\affiliation{ 
 Dipartimento di Fisica Ettore Pancini, Università degli Studi di Napoli Federico II, Via Cinthia, 80126, Napoli, IT.
}%
\author{F. Tafuri}
\affiliation{ 
 Dipartimento di Fisica Ettore Pancini, Università degli Studi di Napoli Federico II, Via Cinthia, 80126, Napoli, IT.
}%

\author{A. Bruno}
\affiliation{%
QuantWare, Elektronicaweg 10, 2628 XG Delft, The Netherlands
}%

\author{D. Massarotti}
\affiliation{Dipartimento di Ingegneria Elettrica e delle Tecnologie dell’Informazione, Università degli Studi di Napoli Federico II, Via Claudio 21, 80125, Napoli, IT 
}%
\date{\today}

\begin{abstract}
We have realized submicron tunnel ferromagnetic Al/AlO$_x$/Al/Ni$_{80}$Fe$_{20}$/Al Josephson junctions (JJs) in Manhattan-style configuration for qubit applications. These junctions have been designed to lie within the energy range of transmons. 
The current-voltage characteristics of these junctions are comparable with those of standard JJs implemented in state-of-the-art transmons, thus confirming the high quality of the devices and marking a significant step toward the realization of the ferrotransmon. Low-frequency characterization confirms that our junctions operate in the quantum phase diffusion limit, as tunnel JJs in conventional transmons with similar characteristic energies. Ultimately, mitigation of quantum phase fluctuations will represent a key for advancing the entire field of superconducting quantum circuit architectures. 

\end{abstract}

\maketitle
  \begin{figure*}[ht]
 \includegraphics[width = 17 cm]{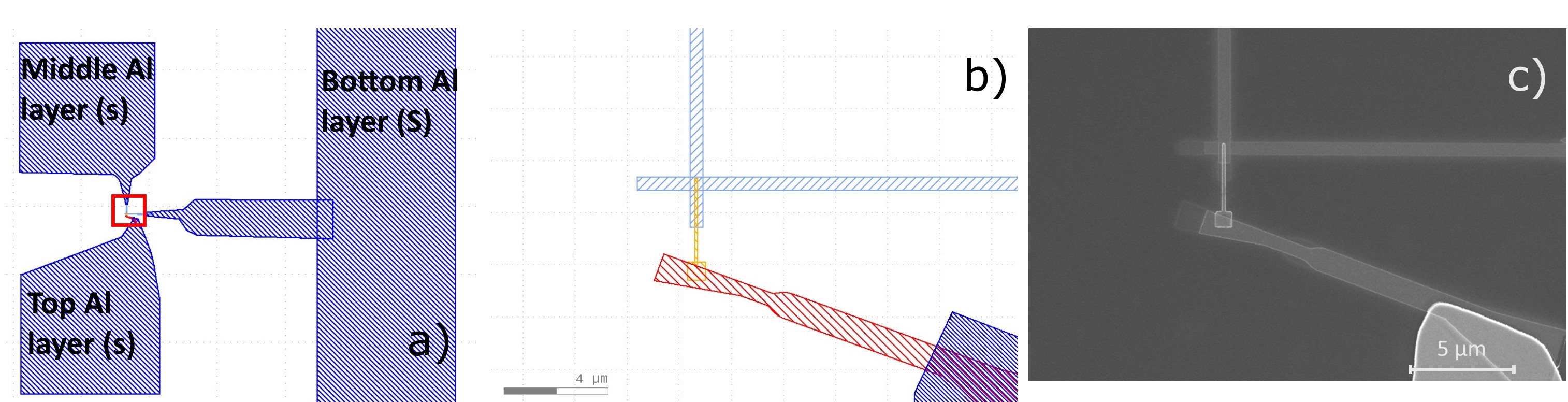}\caption{a) Design of the multilayer Superconductor/ Insulator/ thin superconductor/ Ferromagnet/ Superconductor Josephson junctions (SIsFS JJs). The pads (in blue) allow separate measurements of the SIs, sFS, and multilayer SIsFS junctions. b) Zoom-in of the junction area: the SIs part is shown in light blue, the window for the FS deposition in orange, and the aluminum connection between the junction and the pad in red. c) Scanning Electron Microscope (SEM) image of the fabricated multilayer SIsFS JJ. }   \label{figure:fabrication}
\end{figure*}
Josephson junctions (JJs)  are the core circuit element of superconducting electronics, including quantum coherent devices, arguably the leading technology for quantum computing\cite{Krantz2019,Kjaergaard2019}. Two energy scales govern the physics of these junctions: the Josephson energy \textit{E}$_J$, directly proportional to the superconducting critical current in the absence of fluctuations \textit{I}$_c$, and the charging energy \textit{E}$_C$, which is inversely proportional to the intrinsic junction capacitance \textit{C}$_J$ \cite{tafuri2019}. The unique non-linear properties of JJs allows for the engineering of macroscopic on-chip artificial atoms with transition energies precisely defined by material and circuit design \cite{DiVincenzo,DiCarlo2009,Krantz2019,Kjaergaard2019}. 
Typically, JJs employed in transmon circuits are submicron-sized, fabricated using standard shadow mask techniques like the Dolan bridge \cite{DOlan1977} or Manhattan process \cite{Potts2001,Muthusubramanian2024,Moskalev2023}, and designed to operate in a regime where the Josephson energy is approximately equal to the charging energy of the junction $(E_J \sim E_C)$. To face the sensitivity of small JJs to charge noise, the JJ is shunted by an on-chip capacitor $C_s$. This ensures that the charging energy of the overall circuit $E_{C,\Sigma}$ is much lower than the Josephson energy $E_J$ ($E_J /E_{C,\Sigma} > 50$)\cite{Koch2007}, thus suppressing charge fluctuations. 

Recently,  hybrid paradigms have emerged that enable alternative qubit frequency tuning by integrating JJs employing exotic barriers. For instance, JJs employing semiconducting barriers have been integrated into transmon qubits to achieve tunability using gate voltages (gatemon) rather than magnetic flux \cite{deLange2015,Larsen2015,Ramón2020}.  
A promising approach is based on tunnel magnetic Josephson junctions (MJJs), where a ferromagnetic layer is enclosed in one of the S electrodes\cite{Ahmad2022,Massarotti2023_ferrotransmon,Ahmad2025}. The ferrotransmon concept proposes to exploit the memory properties of tunnel MJJs for qubit frequency control via magnetic field pulses, thus eliminating the need for a static magnetic field during qubit operation. This notion holds substantial promise for scalable superconducting quantum systems driving alternative and energy-efficient cryogenic digital control schemes \cite{Caruso2018,McDermott_2018}. \\
In this work, we have realized submicron Superconductor/ Insulator/ thin superconductor/ Ferromagnet/ Superconductor (SIsFS) JJs, fabricated using the state-of-the-art multilayer procedure by QuantWare for producing transmons used in commercial quantum processors. These junctions have been designed to operate into the series regime of a tunnel SIs and ferromagnetic sFS JJs \cite{Bakurskiy2013}. The former guarantees the low-dissipation and high-quality factors required in highly coherent devices, while the latter enables magnetic switching and alternative tunability schemes of the Josephson energy. We have used advanced fabrication techniques to scale the devices to previously unexplored energy regimes for MJJs, thus marking a fundamental step for the realization of the ferrotransmon. \newline So far, submicrometric MJJs have been mainly realized using Nb technology and focused ion beam processing \cite{Niedzielski2015,Kapran2021}, which complicates the integration into standard superconducting quantum circuits.  Differently from  our previous micron-sized Al-based MJJs \cite{Vettoliere_APL,vettoliere_nanomaterials}, 
these submicron Manhattan-style SIsFS JJs feature a capacitance $C_J$ reduced by more than two orders of magnitude\cite{Ahmad24}. This constitutes the first clear demonstration that MJJs can be scaled to the appropriate energy regime  $(E_{C} \sim E_J)$ and integrated in standard transmon circuits securing the sufficient anharmonicity required for feasible qubit operation. 
Furthermore, the current-voltage (I-V) characteristics, which reveal crucial information on the d.c. junction electrodynamics\cite{Barone1982,tafuri2019,Iansiti1987}, are comparable with those of standard SIS integrated in commercial transmons \cite{Wisne2024}.  Specifically, d.c. measurements have revealed the hallmarks of the quantum phase diffusion (QPD) regime,  expected for $E_{C} \sim E_J$: a finite slope in the superconducting branch, along with a characteristic rounding of the I-V curve close to the switching current $I_{sw}$, even at milliKelvin temperatures 
\cite{Iansiti1987,Iansiti1989}. These findings are thus not strictly tied to the specific layout and nature of the device but rather associated with its energy scales and they can drive further investigation of the effect of quantum phase fluctuations in standard superconducting qubits \cite{Wisne2024}.  
 \newline
The design of the multilayer MJJ, shown in Fig. \ref{figure:fabrication}a, is optimized to enable simultaneous access to three different structures. The SIs junction is probed through the bottom (S) and middle (s) aluminum layers. The sFS junction can be accessed between the middle and top aluminum layers, and the complete SIsFS structure is measured by contacting the bottom and top Al layer pads. The fabrication process of the SIsFS junction involves three distinct lithography steps.  The first step follows the standard Manhattan process used for submicron SIS JJs: electron-beam lithography defines the cross pattern, followed by angled evaporation of Al film and in-situ oxidation to obtain the Al/AlO$_x$/Al trilayer (SIs), represented in light blue in Fig. \ref{figure:fabrication}b. 
 The second lithography step opens a window over the junction area (500 nm x 500 nm) for the deposition of the ferromagnetic material (Ni$_{80}$Fe$_{20}$: Permalloy), after a soft ion milling of the s surface, and the final aluminum layer (S), completing the multilayer JJ — shown in orange in Fig. \ref{figure:fabrication}b. 
 Finally, a third lithography step is used to create the connection between the top aluminum layer of the junction and the bonding pad, represented in red in Fig. \ref{figure:fabrication}b. The final device shown in the Scanning Electron Microscope (SEM) image  in Fig. \ref{figure:fabrication}c has the following layer thicknesses:  Al (20 nm) /AlO$_x$ (2 nm) /Al (25 nm) /Ni$_{80}$Fe$_{20}$ (3 nm)/ Al (70 nm). \\
Two SIsFS JJs, labeled as sample A and B, respectively, have been measured by thermally anchoring the samples to the mixing chamber of a Triton dry dilution refrigerator provided by Oxford instruments, with customized low-noise filters anchored at different temperature stages \cite{Ahmad_22}. The junctions are current-biased with a low-frequency current ramp (approximately 1 Hz) using a waveform generator in series with a shunt resistance, while the voltage across the junction is measured using a differential amplifier.  Magnetic fields in the plane of the junction can be applied using a NbTi coil \cite{Satariano_24}. \\
  \begin{figure}[h]
  \includegraphics[width = 7.5 cm]{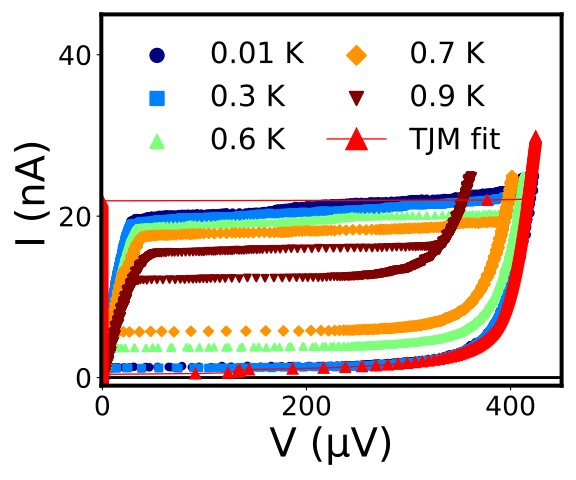} \caption{ Current-voltage (I–V) characteristics of Al (20 nm) /AlO$_x$(2 nm) /Al (25 nm) /Ni$_{80}$Fe$_{20}$ (3 nm)/ Al (70 nm) Josephson junction (Sample A) with a lateral size of 500 nm as a function of the temperature. At low-bias current, the IV characteristic at 10 mK (blue curve) can be fitted with the tunnel junction microscopic (TJM) model (red curve).}   \label{figure:IV}
\end{figure}
The transport properties of SIsFS JJs well fall within the framework of the theory of ferromagnetic JJs proposed by Bakurskiy \textit{et al. }\cite{Bakurskiy2013}.  If the intermediate superconducting layer \(d_{\mathrm{s}}\) is sufficiently larger than the critical thickness \textit{d}\(_{\mathrm{sc}}\), i.e., the minimal thickness of the s layer in a sF bilayer above which superconductivity still exists at a given temperature, the pair potential \(\Delta\)\ in the s layer is close to that of the bulk material. In this scenario, the SIsFS structure can be effectively considered as a series combination of a tunnel SIs JJ and a ferromagnetic sFS JJ \cite{Bakurskiy2013}. For small F-thickness \textit{d}\(_{\mathrm{F}}\), the switching current of the SIs side \( I_{sw}^{SIs}\) is much smaller than that of the sFS side \( I_{sw}^{sFS}\), compare Fig. \ref{figure:IV} with Fig.\ref{figure:appendix}a.  Therefore, since \( I_{sw}^{sFS} \gg  I_{sw}^{SIs}\) across the entire temperature range, the I-V curve of the overall SIsFS device is determined by its SIs part (Fig. \ref{figure:IV}).  Consequently, the temperature dependence of the switching current  \( I_{sw}\)  and of the voltage gap follow the standard behavior of the tunnel SIS JJs\cite{Bakurskiy2013} (Fig. \ref{figure:appendix} b-c). \\
The electrodynamics of the SIsFS JJs in Fig. \ref{figure:IV} is thus dominated solely by the SIs side.  This is a significant finding, as it directly demonstrates that the F layer has no detrimental effect on either the intrinsic phase dynamics or the overall transport regime of the junction for what concerns d. c. characterization, which is a prerequisite for the possible use of tunnel submicron MJJs in transmon circuits.
The I-V curves shown in Fig. \ref{figure:IV} are highly hysteretic and this hysteresis coexists with a slope at low voltage and a characteristic phase diffusion rounding of the switching branch.  This coexistence can be 
justified by taking into account the influence of the embedding circuit through a frequency-dependent quality factor \cite{Kautz1990,Vion1996}. The external environment is characterized by a quality factor $Q_1$, distinct from the intrinsic junction quality factor $Q_0$. Since $Q_1 < Q_0$, the system exhibits a dual damping behavior: at high frequencies near the plasma frequency, $Q_1$ dominates, leading to enhanced damping and
phase diffusion phenomena; at lower frequencies associated with the running state, $Q_0$ prevails, which in turn determines the high value of the subgap resistance \cite{Kautz1990}. \newline
The observed kilo-ohm resistance in the superconducting branch down to dilution temperatures and the rounding of the I-V characteristic at the switching branch provide the first clear indication that the junctions operate in the QPD limit, expected for very small junctions\cite{Iansiti1987,Iansiti1989}. This behavior is fully consistent with what has been observed in standard transmons with similar energy scales\cite{Wisne2024}. While for $E_{C} /E_{J} \ll$1, the phase of the junctions  is localized and can be treated as a semiclassical quantity, in the QPD limit the phase variable is sufficiently delocalized that quantum fluctuations cannot be neglected. For $E_{C} /E_{J}  \sim 1$,  phase delocalization drives multiple escape and retrapping processes. This results in a finite resistance $R_0$ in the superconducting branch and a reduction of the swiching current $ I_{sw}$ relative to the value in absence of quantum fluctuations $I_c$. To account for this scaling, the barrier height of the phase particle potential can be renormalized by introducing the binding energy $ E_{B}=  \frac{\hbar I_{sw}}{2e}$
\cite{Iansiti1989}.  
To verify this hypothesis, we have estimated $E_C$ by fitting the I-V curve in Fig. \ref{figure:IV} with the tunnel junction microscopic (TJM) model. 
The TJM model provides a complete microscopic description of a JJ, using the tunneling-Hamiltonian formalism\cite{Barone1982}. This model can describe the subgap branch and the low-frequency electrodynamics of any JJ that shows tunneling feature transport. The accuracy of the fit is evident from its ability to reproduce the subgap branch of the I-V characteristic.  Following the same procedure as in Ref. [\onlinecite{Ahmad2020}], this fit can provide an estimation of the Stewart-McCumber parameter: $\beta_{C} = \sqrt{Q_0}=\frac{2eI_{sw}C_JR_N^2}{\hbar} \simeq 1$. This value implies a capacitance $C_J$ of the order of  2 fF and a junction charging energy $E_{C}  \sim 40 \  \mu$eV.  We can estimate the bindig energy from the switching current $I_{sw}$: $E_{B} = \hbar I_{sw}/2  e \simeq  40  \ \mu$eV and thus the ratio $ E_{B} /E_{J} \sim1$, thus falling into the QPD regime.  \\
In Fig.~\ref{figure:IV}, the I-V curves as a function of the temperature \textit{T} show a clear dependence
of the finite resistance $R_0$ on the temperature. By linear fitting of the superconducting branch, we estimate the resistance $R_0$ as a function of the temperature $T$ as reported in Fig. \ref{figure:Ro_T}a. Up to 0.5 K, since $E_B > k_BT$, the value of $R_0$ saturates due to freezing out of the thermal fluctuations. Above $T > $ 0.5 K,  $E_B   \leq   k_BT$ and thermal fluctuations mediates phase-diffusion mechanisms, resulting in an exponential behavior of $R_0$ with increasing $T$ \cite{Kautz1990}. 
 \begin{figure}[h]
  \includegraphics[width = 7.5 cm]{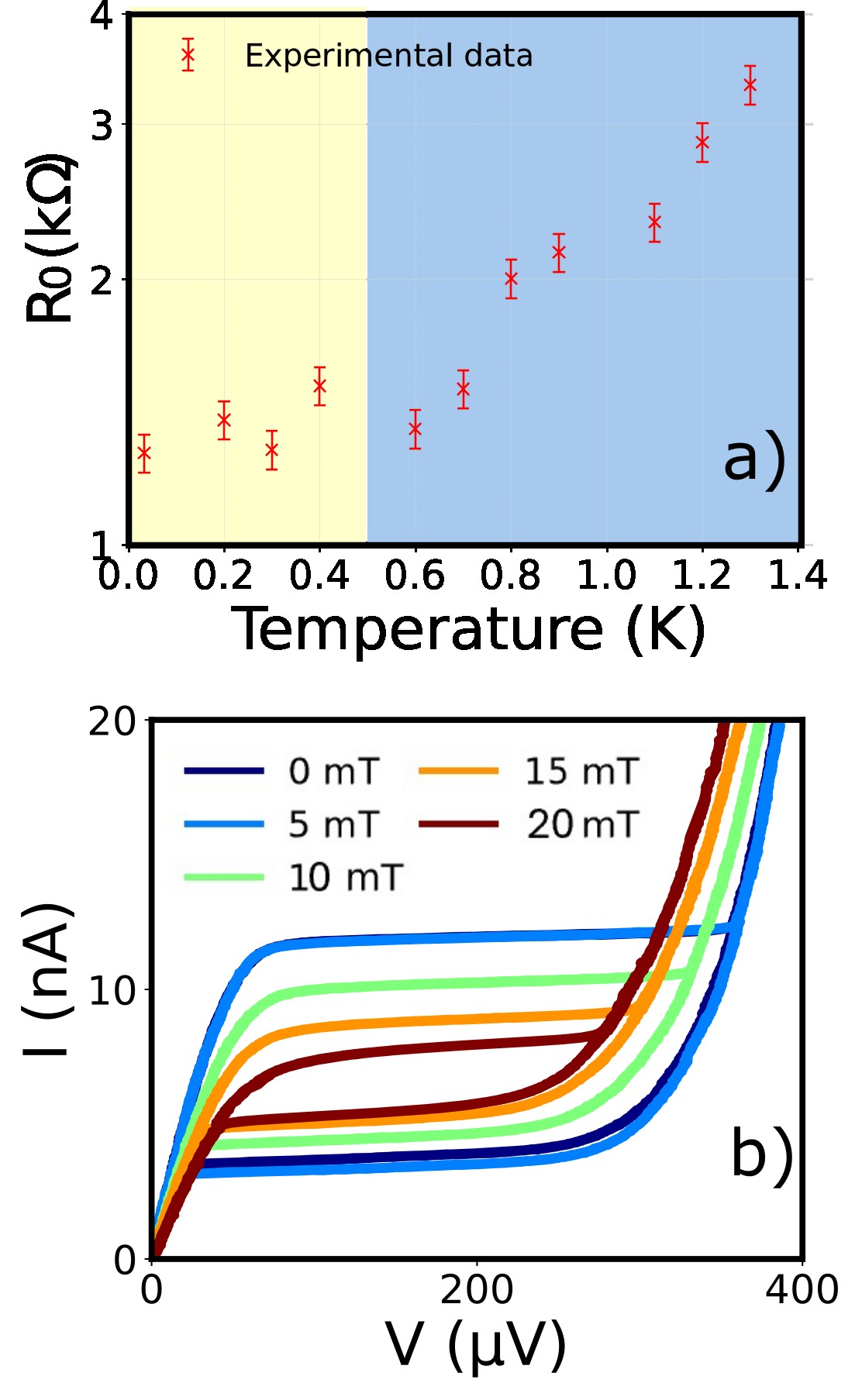} \caption{a) Finite slope of the supercurrent branch of the I-V curves $R_0$ as a function of the temperature $T$, where $R_0$ has been estimated through the linear fitting of the I-V superconducting branch in Figure \ref{figure:IV}. The yellow region indicates the quantum phase diffusion regime, while the light blue one indicates the phase diffusion regime activated by thermal fluctuations. b) Evidence of a resistive branch in the I-V characteristics and its modulation by applying a magnetic field for sample SIsFS B. }\label{figure:Ro_T}
\end{figure}
  \begin{table*}
\caption{\label{table_parameter}Parameters of submicron Superconductor/ Insulator/ thin superconductor/ Ferromagnet/ Superconductor Josephson junctions (SIsFS  JJs) with Al as S electrodes and Ni$_{80}$Fe$_{20}$ as F layer at \textit{T}  = 10 mK: sample, normal resistance \textit{R$_{N}$}, low-bias resistance \textit{R$_{0}$}, switching current \textit{ I$_{sw}$}, binding energy \(E_{B}\), critical current in absence of quantum phase fluctuations \textit{ I$_{c}$},  Josephson energy in absence of fluctuations \(E_{J}\), junction capacitance\textit{ C$_J$},  charging energy $E_C$, the ratio \textit{E}$_{C}$/ \textit{E}$_{B}$ and the ratio \textit{E}$_C/$\textit{E}$_{J}$.}
\begin{ruledtabular}
\begin{tabular}{ccccccccccc}
 Sample &\textit{R}$_N$ (k$\Omega$) & \textit{R}$_{0}$ (k$\Omega$) & \textit{I}$_{sw}$ (nA)& \textit{E}$_B$ ($\mu$eV)&\textit{I}$_{c}$ (nA)& \textit{E}$_J$ ($\mu$eV)& C (fF) 
& \textit{E}$_C$ ($\mu$eV)& \textit{E}$_C/$\textit{E}$_{B}$& \textit{E}$_C/$\textit{E}$_{J}$\\ 
A& 3.2 & 1.3& 20& 40
& 62& 130 
&2
& 40& 1& 0.3\\ 
B& 3.1 & 3.9 & 12& 25& 46& 94 & 2& 40& 1.6 &0.4\\
\end{tabular}
\end{ruledtabular}
\end{table*}
 \begin{figure}[h]
  \includegraphics[width = 8.5 cm]{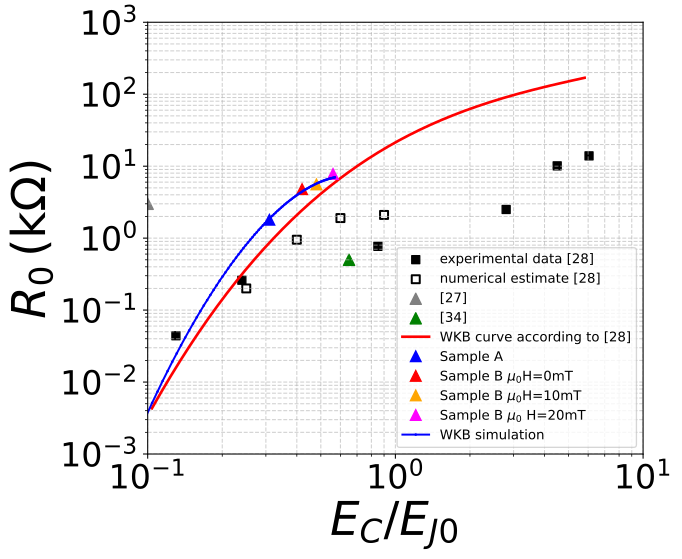} \caption{
 $R_0$ versus the ratio $E_C/E_{J}$. The plot shows a comparison of our experimental data (dots) with theoretical Wentzel–Kramers–Brillouin (WKB) simulations. The red curve, adapted from Ref. [\onlinecite{Iansiti1987}], models quantum tunneling through a potential with a barrier height equal to the Josephson energy ($E_J$), while the blue curve uses the renormalized binding energy ($E_B$). Experimental data from literature are also included for comparison, as detailed in the legend.  }  \label{figure:Ro_H_comparison}
\end{figure}
To provide further compelling evidence of the quantum nature of resistance $R_0$, we have applied a magnetic field to sample SIsFS B at 10 mK, where thermal fluctuations are effectively suppressed $(E_B  \gg k_BT$).  At zero magnetic field, this MJJs exhibits a lower switching current and thus a higher zero-field resistance compared to sample SIsFS A. The magnetic field reduces the switching current \textit{I}$_{sw}$ and thus the binding energy $E_{B}$, which in turn results in an increase of the resistance $R_0$. This finding demonstrates that the observed resistance in the superconducting branch is a fundamental feature due to the tuning of the binding energy, rather than a spurious effect due to the fabrication process or electromagnetic noise. \\
In Fig. \ref{figure:Ro_H_comparison}, we compare our measured resistance values $R_0$ with a Wentzel–Kramers–Brillouin (WKB) simulation from Ref. [\onlinecite{Iansiti1987}]. This simulation models quantum tunneling of the phase through a potential well with a barrier height corresponding to the Josephson energy $E_{J}$ in the absence of quantum phase fluctuations. 
The value of $E_{J}$ for our samples has been estimated using the relation between the binding energy $E_B$ and the ratio $x = E_c/E_{J}$ in the QPD limit\cite{Iansiti1987}:
\begin{equation}
\label{binding_energy}
     E_B = \frac{\hbar I_{sw}}{2e} = E_{J}\, 2x \left[ \left( 1 + \frac{1}{8x^2} \right)^{1/2} - 1 \right].
\end{equation}
By substituting the parameters reported in Table~\ref{table_parameter} into Eq.~\ref{binding_energy}, we obtain a value of the Josephson energy $E_{J} = 130 \,\mu\text{eV}$ for Sample~A and $94 \,\mu\text{eV}$ for Sample~B, respectively. 
Moreover, the scaling of the Josephson energy results in a reduction of the switching current with respect to the critical current fluctuation-free value $I_{c} =\frac{2e}{\hbar}E_{J} $. This reduction of $70\%$ is consistent with previous reports\cite{Wisne2024,Stornaiuolo2013}.
The WKB simulation from Ref.~[\onlinecite{Iansiti1987}] (red curve in Fig. \ref{figure:Ro_H_comparison} ) shows good agreement with our experimental data.  An even better consistency is obtained when considering a WKB simulation, where the barrier height corresponds to the binding energy $E_B$ (blue curve in Fig. \ref{figure:Ro_H_comparison}).\\
In summary, we have realized the first generation of high-quality submicrometric tunnel MJJs using standard fabrication techniques for superconducting computation quantum architectures. The SIsFS layout allows us to preserve the d.c. transport properties of the tunnel submicrometer SIS side, thus ensuring low levels of dissipation and energy scales fully compatible with the transmon regime. Results are comparable to those of standard Al-based JJs employed in conventional transmons, further supported by the observation that our MJJs operate within the quantum phase diffusion limit.  Moreover,  in submicron MJJs in the appropriate energy range, the Josephson energy can no longer be directly correlated with the switching current $I_{sw}$. 
Comparative studies on non-conventional JJs, as MJJs, may inspire methods to mitigate and better control quantum phase fluctuations. Scaling of the SIsFS JJ's would allow for their integration  in standard transmon circuits.  However, these SIsFS JJs require magnetic field pulses above 30 mT to be tuned at zero-static magnetic field \cite{Parlato2020,Vettoliere_APL}, which are incompatible with efficient on-chip control\cite{Ahmad2025}. Current work focuses on replacing the actual ferromagnetic barrier with a softer ferromagnetic layer (e.g., Nb-doped permalloy \cite{Birge_review,Baek2014}), as the ferromagnetic material can be integrated \textit{ex-situ} without significantly affecting the junction electrodynamics. This approach will make the technology viable for an alternative tuning of the qubit frequency into quantum circuits \cite{Ahmad2022}.  Further investigations adopting noise-detection protocols in RF environment described in Ref. [\onlinecite{Ahmad2022, Ahmad2023}] will enable us to effectively discriminate noise arising from the F barrier from other dominant sources of noise in standard transmon architecture, e.g., Purcell effect\cite{Houck2008}, quasi-particle\cite{serniak2018} and flux-noise fluctuations\cite{Hutchings2017}. 

\setcounter{figure}{0}
\makeatletter 
\renewcommand{\thefigure}{A\@arabic\c@figure}
\makeatother

\begin{figure*}[h ]
  \includegraphics[width = 18cm]{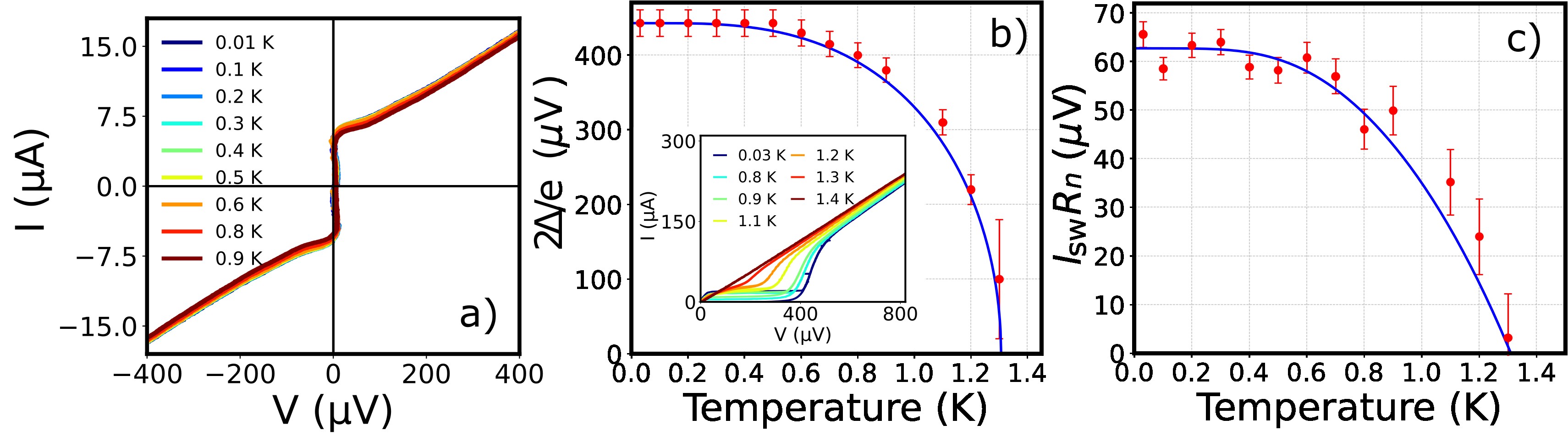}\caption{a) Measurements of the I–V characteristics for the sFS trilayer within the SIsFS multilayer JJs as a function of the temperature \textit{T}. b) The inset shows the temperature dependence of the I–V characteristics for SIsFS JJ A. From these curves,  the values of the superconducting voltage gap \(2 \Delta/e\) [panel b)] and the characteristic voltage \(I_{sw} R_{N}\) [panel c)] have been determined. In both panels b) and c), the experimental data (red points) have been fitted (red curves) by using the Bardeen-Cooper-Schrieffer (BCS) equation for the superconducting gap [panel b)] and the  Ambegaokar-Baratoff (AB) relation for the \(I_{sw} R_{N}\) product [panel c)]. 
 }   \label{figure:appendix}
\end{figure*}
\appendix*
\section{Serial transport regime}

Measurements as a function of temperature up to 1.4 K are reported in Fig. \ref{figure:appendix}a for the sFS side of the SIsFS A device.  From each IV curve of the SIsFS JJ A in the inset of Fig. \ref{figure:appendix}b, the gap voltage \( 2 \Delta / e\) (Fig. \ref{figure:appendix}b) and the \(I_{sw}R_{N}\) product (Fig. \ref{figure:appendix}c) have been extracted. The experimental temperature dependence \( 2 \Delta / e\) follows the  Bardeen-Cooper-Schrieffer (BCS)  approximation in the weak-coupling limit: \(\frac{2 \Delta(T)} {e}  = \frac{2 \Delta_{0}}{e} \tanh \left ( 1.74 \sqrt{1 - \frac{T}{T_{c}}} \right )\) \cite{Barone1982}. The fit reported in Fig. \ref{figure:appendix}b provides the zero temperature $ V_{gap} = 440 \pm 20 \mu \textrm{eV}$ and $T_c = 1.3 \pm 0.1$. The \(I_{sw}R_{N}(T)\) curves follow the Ambegaokar-Baratoff (AB)  relation \cite{AB_theory}:  \(I_{sw}R_{N}(T) = A \frac{\pi}{2e} \Delta(T)\tanh \left ( \frac{\Delta(T)}{2k_{B}T} \right )\). In this case, we have introduced a coefficient \(A\) to taken into account the suppression of the critical current with respect to the ideal case, which is equal to  $0.18 \pm 0.02$.

\section*{Data Availability Statement}
The data that support the findings of this study are available from the corresponding author upon reasonable request.
\section*{Conflict of Interest statement}
The authors have no conflicts to disclose. 
\begin{acknowledgments}
This work has been supported by the Pathfinder EIC 2023 project "FERROMON-Ferrotransmons and Ferrogatemons for Scalable Superconducting Quantum Computers", the PNRR MUR project PE0000023-NQSTI, the PNRR MUR project CN-00000013-ICSC, 
\end{acknowledgments}

\bibliography{aipsamp}

@book{tafuri2019,
  title={Fundamentals and frontiers of the Josephson effect},
  author={Tafuri, Francesco},
  volume={286},
  year={2019},
  publisher={Springer Nature}
}

@article{Krantz2019,
	author = {Krantz,P.  and Kjaergaard,M.  and Yan,F.  and Orlando,T. P.  and Gustavsson,S.  and Oliver,W. D. },
	title = {A quantum engineer's guide to superconducting qubits},
	journal = {Applied Physics Reviews},
	volume = {6},
	number = {2},
	pages = {021318},
	year = {2019},
	doi = {10.1063/1.5089550},
	
	URL = { 
	https://doi.org/10.1063/1.5089550 },
	eprint = { https://doi.org/10.1063/1.5089550}	
}

@article{Vion1996,
  title = {Thermal Activation above a Dissipation Barrier: Switching of a Small Josephson Junction},
  author = {Vion, D. and G\"otz, M. and Joyez, P. and Esteve, D. and Devoret, M. H.},
  journal = {Phys. Rev. Lett.},
  volume = {77},
  issue = {16},
  pages = {3435--3438},
  numpages = {0},
  year = {1996},
  month = {Oct},
  publisher = {American Physical Society},
  doi = {10.1103/PhysRevLett.77.3435},
  url = {https://link.aps.org/doi/10.1103/PhysRevLett.77.3435}
}

@article{DiCarlo2009,
	author={DiCarlo, L.
	and Chow, J. M.
	and Gambetta, J. M.
	and Bishop, Lev S.
	and Johnson, B. R.
	and Schuster, D. I.
	and Majer, J.
	and Blais, A.
	and Frunzio, L.
	and Girvin, S. M.
	and Schoelkopf, R. J.},
	title={Demonstration of two-qubit algorithms with a superconducting quantum processor},
	journal={Nature},
	year={2009},
	month={Jul},
	day={01},
	volume={460},
	number={7252},
	pages={240-244},
	issn={1476-4687},
	doi={10.1038/nature08121},
	url={https://doi.org/10.1038/nature08121}
}

@article{Kjaergaard2019,
	author = {Kjaergaard, Morten and Schwartz, Mollie E. and Braumüller, Jochen and Krantz, Philip and Wang, Joel I.-J. and Gustavsson, Simon and Oliver, William D.},
	title = {Superconducting Qubits: Current State of Play},
	journal = {Annual Review of Condensed Matter Physics},
	volume = {11},
	number = {1},
	pages = {369-395},
	year = {2020},
	doi = {10.1146/annurev-conmatphys-031119-050605},
	
	URL = {	https://doi.org/10.1146/annurev-conmatphys-031119-050605},
	eprint = {https://doi.org/10.1146/annurev-conmatphys-031119-050605}	
}

@article{DOlan1977,
    author = {Dolan, G. J.},
    title = {Offset masks for lift‐off photoprocessing},
    journal = {Appl. Phys. Lett.},
    volume = {31},
    number = {5},
    pages = {337-339},
    year = {1977},
    month = {09},    
    issn = {0003-6951},
    doi = {10.1063/1.89690},
    url = {https://doi.org/10.1063/1.89690}
}

@article{Potts2001,
    author = {Potts, A. and Routley, P. R. and Parker, G. J. and Baumberg, J. J. and de Groot, P. A. J.},
    title = {Novel fabrication methods for submicrometer Josephson junction qubits},
    journal = {Journal of Materials Science: Materials in Electronics},
    volume = {12},
    number = {4},
    pages = {289-293},
    year = {2001},
    doi = {10.1023/A:1011279908265}
}

@article{Muthusubramanian2024,
    author = {Muthusubramanian, N. and Finkel,M. and Duivestein,P. and Zachariadis, C. and van der Meer, Sean L M and Veen,Hendrik M and Beekman, M. W. and  Stavenga, T. and   Bruno, Alessandro and  DiCarlo, Leonardo},
    title = {Wafer-scale uniformity of Dolan-bridge and bridgeless Manhattan-style Josephson junctions for superconducting quantum processors},
    journal = {Quantum Science and Technology},
    volume = {9},
    number = {2},
    pages = {025006},
    year = {2024},
    doi = {10.1088/2058-9565/ad199c}
}

@article{Moskalev2023,
    author = {Moskalev, D. O. and Zikiy, E. V. and Pishchimova, A. A. and Ezenkova, D. A. and Smirnov, N. S. and Ivanov, A. I. and Korshakov, N. D. and Rodionov, I. A.O.},
    title = {Optimization of shadow evaporation and oxidation for reproducible quantum Josephson junction circuits},
    journal = {Scientific Reports},
    volume = {13},
    number = {1},
    pages = {4174},
    year = {2023},
    doi = {10.1038/s41598-023-31003-1}
}

@article{Ramón2020,
    author = {Aguado, Ramón},
    title = {A perspective on semiconductor-based superconducting qubits},
    journal = {Applied Physics Letters},
    volume = {117},
    number = {24},
    pages = {240501},
    year = {2020},
    month = {12},
    doi = {10.1063/5.0024124}
}

@article{Larsen2015,
  title = {Realization of Microwave Quantum Circuits Using Hybrid Superconducting-Semiconducting Nanowire Josephson Elements},
  author = {de Lange, G. and van Heck, B. and Bruno, A. and van Woerkom, D. J. and Geresdi, A. and Plissard, S. R. and Bakkers, E. P. A. M. and Akhmerov, A. R. and DiCarlo, L.},
  journal = {Phys. Rev. Lett.},
  volume = {115},
  issue = {12},
  pages = {127002},
  numpages = {5},
  year = {2015},
  month = {Sep},
  publisher = {American Physical Society},
  doi = {10.1103/PhysRevLett.115.127002},
  url = {https://link.aps.org/doi/10.1103/PhysRevLett.115.127002}
}

@article{deLange2015,
  title = {Realization of Microwave Quantum Circuits Using Hybrid Superconducting-Semiconducting Nanowire Josephson Elements},
  author = {de Lange, G. and van Heck, B. and Bruno, A. and van Woerkom, D. J. and Geresdi, A. and Plissard, S. R. and Bakkers, E. P. A. M. and Akhmerov, A. R. and DiCarlo, L.},
  journal = {Phys. Rev. Lett.},
  volume = {115},
  issue = {12},
  pages = {127002},
  numpages = {5},
  year = {2015},
  month = {Sep},
  publisher = {American Physical Society},
  doi = {10.1103/PhysRevLett.115.127002},
  url = {https://link.aps.org/doi/10.1103/PhysRevLett.115.127002}
}

@article{McDermott_2018,
doi = {10.1088/2058-9565/aaa3a0},
year = {2018},
volume = {3},
number = {2},
pages = {024004},
author = { McDermott, R. and Vavilov, M. G.  and Plourde, B. L. T. and Wilhelm,  F. K . and Liebermann,P. J. and Mukhanov, O. A.  and Ohki,  T. A. },
title = {Quantum–classical interface based on single flux quantum digital logic},
journal = {Quantum Sci. Technol.}

}

@article{Massarotti2023_ferrotransmon,
    author = {Massarotti, D. and Ahmad, H. G. and Satariano, R. and Ferraiuolo, R. and Di Palma, L. and Mastrovito, P. and Serpico, G. and Levochkina, A. and Caruso, R. and Miano, A. and Arzeo, M. and Ausanio, G. and Granata, C. and Lucignano, P. and Montemurro, D. and Parlato, L. and Vettoliere, A. and Fazio, R. and Mukhanov, O. and Pepe, G. P. and Tafuri, F.},
    title = {A feasible path for the use of ferromagnetic josephson junctions in quantum circuits: The ferro-transmon},
    journal = {Low Temperature Physics},
    volume = {49},
    number = {7},
    pages = {794-802},
    year = {2023},
    month = {07},
    issn = {1063-777X},
    doi = {10.1063/10.0019690},
    url = {https://doi.org/10.1063/10.0019690}
}

@ARTICLE{Ahmad2025,
  author={Ahmad, Halima Giovanna and Ferraiuolo, Raffaella and Serpico, Giuseppe and Satariano, Roberta and Levochkina, Anna and Vettoliere, Antonio and Granata, Carmine and Montemurro, Domenico and Esposito, Martina and Ausanio, Giovanni and Parlato, Loredana and Pepe, Giovanni Piero and Bruno, Alessandro and Tafuri, Francesco and Massarotti, Davide},
  journal={IEEE Transactions on Applied Superconductivity}, 
  title={Towards Novel Tunability Schemes for Hybrid Ferromagnetic Transmon Qubits}, 
  year={2025},
  volume={35},
  number={5},
  pages={1-7},
 doi={10.1109/TASC.2025.3535674}}

@Article{vettoliere_nanomaterials,
AUTHOR = {Vettoliere, A. and Satariano, R. and Ferraiuolo, R. and Di Palma, L. and Ahmad, H. G. and Ausanio, G. and Pepe, G. P. and Tafuri, F. and Massarotti, D. and Montemurro, D. and Granata, C. and Parlato, .},
TITLE = {High-Quality Ferromagnetic Josephson Junctions Based on Aluminum Electrodes},
JOURNAL = {Nanomaterials},
VOLUME = {12},
YEAR = {2022},
NUMBER = {23},
ARTICLE-NUMBER = {4155},
DOI = {10.3390/nano12234155}
}

@article{Vettoliere_APL,
author = {Vettoliere,A.  and Satariano,R.  and Ferraiuolo,R.  and Di Palma,L.  and Ahmad,H. G.  and Ausanio,G.  and Pepe,G. P.  and Tafuri,F.  and Montemurro,D.  and Granata,C.  and Parlato,L.  and Massarotti,D. },
title = {Aluminum-ferromagnetic Josephson tunnel junctions for high quality magnetic switching devices},
journal = {Appl. Phys. Lett.},
volume = {120},
number = {26},
pages = {262601},
year = {2022},
doi = {10.1063/5.0101686}

}

@article{serniak2018,
  title = {Hot Nonequilibrium Quasiparticles in Transmon Qubits},
  author = {Serniak, K. and Hays, M. and de Lange, G. and Diamond, S. and Shankar, S. and Burkhart, L. D. and Frunzio, L. and Houzet, M. and Devoret, M. H.},
  journal = {Phys. Rev. Lett.},
  volume = {121},
  issue = {15},
  pages = {157701},
  numpages = {6},
  year = {2018},
  month = {Oct},
  publisher = {American Physical Society},
  doi = {10.1103/PhysRevLett.121.157701},
  url = {https://link.aps.org/doi/10.1103/PhysRevLett.121.157701}
}

@article{Ahmad24,
    author = {Ahmad, H. G. and Satariano, R. and Ferraiuolo, R. and Vettoliere, A. and Granata, C. and Montemurro, D. and Ausanio, G. and Parlato, L. and Pepe, G. P. and Tafuri, F. and Massarotti, D.},
    title = {Phase dynamics of tunnel Al-based ferromagnetic Josephson junctions},
    journal = {Applied Physics Letters},
    volume = {124},
    number = {23},
    pages = {232601},
    year = {2024},
    month = {06},
    doi = {10.1063/5.0211006}
}

@article{Kapran2021,
  title = {Crossover between short- and long-range proximity effects in superconductor/ferromagnet/superconductor junctions with Ni-based ferromagnets},
  author = {Kapran, O. M. and Golod, T. and Iovan, A. and Sidorenko, A. S. and Golubov, A. A. and Krasnov, V. M.},
  journal = {Phys. Rev. B},
  volume = {103},
  issue = {9},
  pages = {094509},
  numpages = {13},
  year = {2021},  
  publisher = {American Physical Society},
  doi = {10.1103/PhysRevB.103.094509},
  url = {https://link.aps.org/doi/10.1103/PhysRevB.103.094509}
}

@article{Wisne2024,
  title = {Transport signatures of phase fluctuations in superconducting qubits},
  author = {Wisne, M. and Deng,  Y. and Cansizoglu,H. and Kopas, C. and  Mutus, J. Y. and Chandrasekhar, V.},
  journal = {Mater. Quantum. Technol. },
  volume = {4},
  issue = {4},
  pages = {046001},
  year = {2024},  
  doi = {DOI 10.1088/2633-4356/ad9a4}
}

@article{DiVincenzo,
author = {DiVincenzo, David P.},
title = {The Physical Implementation of Quantum Computation},
journal = {Fortschritte der Physik},
volume = {48},
number = {9-11},
pages = {771-783},
doi = {https://doi.org/10.1002/1521-3978(200009)48:9/11<771::AID-PROP771>3.0.CO;2-E},
year = {2000}
}

@article{AB_theory,
  title = {Tunneling Between Superconductors},
  author = {Ambegaokar, V. and Baratoff,A. },
  journal = { Phys. Rev. Lett.},
  volume = {10},
  issue = {11},
  pages = { 486 - 489}, 
  year = {1963}
  
 
}

@article{Bakurskiy2013,
	author = {Bakurskiy,S. V.  and Klenov,N. V.  and Soloviev,I. I.  and Bol'ginov,V. V.  and Ryazanov,V. V.  and Vernik,I. V.  and Mukhanov,O. A.  and Kupriyanov,M. Yu.  and Golubov,A. A. },
	title = {Theoretical model of superconducting spintronic {SIsFS} devices},
	journal = {Applied Physics Letters},
	volume = {102},
	number = {19},
	pages = {192603},
	year = {2013},
	doi = {10.1063/1.4805032},
	
	URL = { 
	https://doi.org/10.1063/1.4805032
	
	},
	eprint = { 
	https://doi.org/10.1063/1.4805032
	
	}
	
}

@article{Birge_review,
    author = {Birge, Norman O. and Satchell, Nathan},
    title = {Ferromagnetic materials for $\pi$ Josephson junctions},
    journal = {APL Materials},
    volume = {12},
    number = {4},
    pages = {041105},
    year = {2024},
    month = {04},
    issn = {2166-532X},
    doi = {10.1063/5.0195229},
    url = {https://doi.org/10.1063/5.0195229}
}

@article{Niedzielski2015,
	doi = {10.1088/0953-2048/28/8/085012},
	url = {https://doi.org/10.1088/0953-2048/28/8/085012},
	year = {2015},
	month = {jul},
	publisher = {{IOP} Publishing},
	volume = {28},
	number = {8},
	pages = {085012},
	author = {Bethany M Niedzielski and E C Gingrich and Reza Loloee and W P Pratt and Norman O Birge},
	title = {S/F/S Josephson junctions with single-domain ferromagnets for memory applications},
	journal = {Superconductor Science and Technology}
}

@article{Koch2007,
	title = {Charge-insensitive qubit design derived from the {Cooper} pair box},
	author = {Koch, J. and Yu, Terri M. and Gambetta, Jay and Houck, A. A. and Schuster, D. I. and Majer, J. and Blais, A. and Devoret, M. H. and Girvin, S. M. and Schoelkopf, R. J.},
	journal = {Phys. Rev. A},
	volume = {76},
	issue = {4},
	pages = {042319},
	numpages = {19},
	year = {2007},
	month = {Oct},
	publisher = {American Physical Society},
	doi = {10.1103/PhysRevA.76.042319},
	url = {https://link.aps.org/doi/10.1103/PhysRevA.76.042319}
}

@article{Caruso2018,
	author = {Caruso, R. and Massarotti, D. and V. V. and Ben-Hamida, A. and Karelina, N.L. and Miano, A. and Vernik, I. and Tafuri, F. and Ryazanov, V. and Mukhanov, O. and Pepe, G. P.},
	year = {2018},
	month = {04},
	pages = {133901},
	title = {{RF} assisted switching in magnetic {Josephson} junctions},
	volume = {123},
	journal = {Journal of Applied Physics},
	doi = {10.1063/1.5018854}
}

@article{Ahmad2020,
	title = {Electrodynamics of Highly Spin-Polarized Tunnel {Josephson} Junctions},
	author = {Ahmad, H.G. and Caruso, R. and Pal, A. and Rotoli, G. and Pepe, G.P. and Blamire, M.G. and Tafuri, F. and Massarotti, D.},
	journal = {Phys. Rev. Applied},
	volume = {13},
	issue = {1},
	pages = {014017},
	numpages = {10},
	year = {2020},
	month = {Jan},
	publisher = {American Physical Society},
	doi = {10.1103/PhysRevApplied.13.014017},
	url = {https://link.aps.org/doi/10.1103/PhysRevApplied.13.014017}
}

@article{Ahmad2022,
  title = {Hybrid ferromagnetic transmon qubit: Circuit design, feasibility, and detection protocols for magnetic fluctuations},
  author = {Ahmad, Halima Giovanna and Brosco, Valentina and Miano, Alessandro and Di Palma, Luigi and Arzeo, Marco and Montemurro, Domenico and Lucignano, Procolo and Pepe, Giovanni Piero and Tafuri, Francesco and Fazio, Rosario and Massarotti, Davide},
  journal = {Phys. Rev. B},
  volume = {105},
  issue = {21},
  pages = {214522},
  numpages = {14},
  year = {2022},
  month = {Jun},
  publisher = {American Physical Society},
  doi = {10.1103/PhysRevB.105.214522},
  url = {https://link.aps.org/doi/10.1103/PhysRevB.105.214522}
}

@article{Satariano_24,
  title = {Nanoscale spin ordering and spin screening effects in tunnel ferromagnetic Josephson junctions.},
  author = { Satariano, Roberta
and Volkov, Anatoly Fjodorovich
and Ahmad, Halima Giovanna
and Di Palma, Luigi
and Ferraiuolo, Raffaella
and Vettoliere, Antonio
and Granata, Carmine
and Montemurro, Domenico
and Parlato, Loredana
and Pepe, Giovanni Piero
and Tafuri, Francesco
and Ausanio, Giovanni
and Massarotti, Davide},
  journal = {Commun. Mater.},
  volume = {5},
  issue = {1},
  pages= {67},
  year = {2024},
 doi = {10.1038/s43246-024-00497-1}
  
}

@article{Parlato2020,
	author = {Parlato,Loredana  and Caruso,Roberta  and Vettoliere,Antonio  and Satariano,Roberta  and Ahmad,Halima Giovanna  and Miano,Alessandro  and Montemurro,Domenico  and Salvoni,Daniela  and Ausanio,Giovanni  and Tafuri,Francesco  and Pepe,Giovanni Piero  and Massarotti,Davide  and Granata,Carmine },
	title = {Characterization of scalable {Josephson} memory element containing a strong ferromagnet},
	journal = {Journal of Applied Physics},
	volume = {127},
	number = {19},
	pages = {193901},
	year = {2020},
	doi = {10.1063/5.0004554},
	
	URL = { 
	https://doi.org/10.1063/5.0004554
	
	},
	eprint = { 
	https://doi.org/10.1063/5.0004554
	
	}
	
}

@article{Iansiti1989,
	title = {Charging effects and quantum properties of small superconducting tunnel junctions},
	author = {Iansiti, M. and Tinkham, M. and Johnson, A. T. and Smith, Walter F. and Lobb, C. J.},
	journal = {Phys. Rev. B},
	volume = {39},
	issue = {10},
	pages = {6465--6484},
	numpages = {0},
	year = {1989},
	month = {Apr},
	publisher = {American Physical Society},
	doi = {10.1103/PhysRevB.39.6465},
	url = {https://link.aps.org/doi/10.1103/PhysRevB.39.6465}
}

@article{Ahmad_22,
  title = {Coexistence and tuning of spin-singlet and triplet transport in spin-filter {J}osephson junctions},
  author = {Ahmad, H. G. and Minutillo, M. and Capecelatro, R. and Pal, A. and Caruso, R. and Passarelli, G. and Blamire, M. G. and Tafuri, F. and Lucignano, P. and  Massarotti, D. },
  journal = {Commun. Phys.},
  volume = {5},
  issue = {2},
  
  year = {2022}  
}

@BOOK{Barone1982,
	author={A. Barone and G. Paternò},
	title={Physics and Application of the {Josephson} Effect},
	year={1982},
	publisher={John Wiley and Sons}}

@article{Iansiti1987,
	title = {Charging energy and phase delocalization in single very small {Josephson} tunnel junctions},
	author = {Iansiti, M. and Johnson, A. T. and Smith, Walter F. and Rogalla, H. and Lobb, C. J. and Tinkham, M.},
	journal = {Phys. Rev. Lett.},
	volume = {59},
	issue = {4},
	pages = {489--492},
	numpages = {0},
	year = {1987},
	month = {Jul},
	publisher = {American Physical Society},
	doi = {10.1103/PhysRevLett.59.489},
	url = {https://link.aps.org/doi/10.1103/PhysRevLett.59.489}
}

@Article{Baek2014,
	author={Baek, Burm and Rippard, William H. 	and Benz, Samuel P. and Russek, Stephen E. and Dresselhaus, Paul D.},
	title={Hybrid superconducting magnetic memory device using competing order parameters},
	journal={Nature Communications},
	year={2014},
	month={May},
	day={28},
	volume={5},
	number={1},
	pages={3888},
	doi={10.1038/ncomms4888}
}

@article{Kautz1990,
	title = {Noise-affected {IV} curves in small hysteretic {Josephson} junctions},
	author = {Kautz, R. L. and Martinis, J. M.},
	journal = {Phys. Rev. B},
	volume = {42},
	issue = {16},
	pages = {9903--9937},
	numpages = {0},
	year = {1990},
	month = {Dec},
	publisher = {American Physical Society},
	doi = {10.1103/PhysRevB.42.9903},
	url = {https://link.aps.org/doi/10.1103/PhysRevB.42.9903}
}

@article{Stornaiuolo2013,
	title = {Resolving the effects of frequency-dependent damping and quantum phase diffusion in {YBCO Josephson} junctions},
	author = {Stornaiuolo, D. and Rotoli, G. and Massarotti, D. and Carillo, F. and Longobardi, L. and Beltram, F. and Tafuri, F.},
	journal = {Phys. Rev. B},
	volume = {87},
	issue = {13},
	pages = {134517},
	numpages = {7},
	year = {2013},
	month = {Apr},
	publisher = {American Physical Society},
	doi = {10.1103/PhysRevB.87.134517},
	url = {https://link.aps.org/doi/10.1103/PhysRevB.87.134517}
}

@article{Hutchings2017,
	title = {Tunable Superconducting Qubits with Flux-Independent Coherence},
	author = {Hutchings, M. D. and Hertzberg, J. B. and Liu, Y. and Bronn, N. T. and Keefe, G. A. and Brink, Markus and Chow, Jerry M. and Plourde, B. L. T.},
	journal = {Phys. Rev. Applied},
	volume = {8},
	issue = {4},
	pages = {044003},
	numpages = {13},
	year = {2017},
	month = {Oct},
	publisher = {American Physical Society},
	doi = {10.1103/PhysRevApplied.8.044003},
	url = {https://link.aps.org/doi/10.1103/PhysRevApplied.8.044003}
}

@article{Houck2008,
  title = {Controlling the Spontaneous Emission of a Superconducting Transmon Qubit},
  author = {Houck, A. A. and Schreier, J. A. and Johnson, B. R. and Chow, J. M. and Koch, Jens and Gambetta, J. M. and Schuster, D. I. and Frunzio, L. and Devoret, M. H. and Girvin, S. M. and Schoelkopf, R. J.},
  journal = {Phys. Rev. Lett.},
  volume = {101},
  issue = {8},
  pages = {080502},
  numpages = {4},
  year = {2008},
  month = {Aug},
  publisher = {American Physical Society},
  doi = {10.1103/PhysRevLett.101.080502},
  url = {https://link.aps.org/doi/10.1103/PhysRevLett.101.080502}
}

@ARTICLE{Ahmad2023,
  author={Ahmad, H. G. and Brosco, V. and Miano, A. and Di Palma, L. and Arzeo, M. and Satariano, R. and Ferraiuolo, R. and Lucignano, P. and Vettoliere, A. and Granata, C. and Parlato, L. and Ausanio, G. and Montemurro, D. and Pepe, G. P. and Fazio, R. and Tafuri, F. and Massarotti, D.},
  journal={IEEE Transactions on Applied Superconductivity}, 
  title={Competition of Quasiparticles and Magnetization Noise in Hybrid Ferromagnetic Transmon Qubits}, 
  year={2023},
  volume={33},
  number={5},
  pages={1-6},
  doi={10.1109/TASC.2023.3243197}}
\end{document}